\documentclass[10pt,twocolumn]{IEEEtran}
\usepackage{graphicx}
\usepackage{amsmath}
\usepackage{amssymb}
\usepackage[caption=false]{subfig}
\usepackage[noadjust]{cite}
\usepackage{float}
\usepackage{algorithm}
\usepackage{bibentry}
\usepackage{balance}
\usepackage{algorithm}
\usepackage{algorithmicx}
\usepackage{algpseudocode}
\usepackage{xcolor}

\begin{document}

\title{SLNR Based Precoding for One-Bit Quantized Massive MIMO in mmWave Communications}

\author{Yavuz Yap{\i}c{\i}$^*$, Sung Joon Maeng$^*$, \.{I}smail G\"{u}ven\c{c}$^*$, Huaiyu Dai$^*$, and Arupjyoti Bhuyan$^\dagger$\\
$^*$Department of Electrical and Computer Engineering, North Carolina State University, Raleigh, NC\\
$^\dagger$Idaho National Laboratory, Idaho Falls, ID\\
{\tt \{yyapici,smaeng,iguvenc,hdai\}@ncsu.edu, arupjyoti.bhuyan@inl.gov}
\thanks{Work supported through the INL Laboratory Directed Research \& Development (LDRD) Program under DOE Idaho Operations Office Contract DE-AC07-05ID14517.}}%

\maketitle

\begin{abstract}
Massive multiple-input multiple-output (MIMO) is a key technology for 5G wireless communications with a promise of significant capacity increase. The use of low-resolution data converters is crucial for massive MIMO to make the overall transmission as cost- and energy-efficient as possible. In this work, we consider a downlink millimeter-wave (mmWave) transmission scenario, where multiple users are served simultaneously by massive MIMO with one-bit digital-to-analog (D/A) converters. In particular, we propose a novel precoder design based on signal-to-leakage-plus-noise ratio (SLNR), which minimizes energy leakage into undesired users while taking into account impairments due to nonlinear one-bit quantization. We show that well-known regularized zero-forcing (RZF) precoder is a particular version of the proposed SLNR-based precoder, which is obtained when quantization impairments are totally ignored. Numerical results underscore significant performance improvements along with the proposed SLNR-based precoder as compared to either RZF or zero-forcing (ZF) precoders.
\end{abstract}


\section{Introduction}\label{sec:intro}

The unprecedented demand on high-speed data rate for next-generation wireless systems makes massive MIMO a key technology for 5G communications. The use of massive number of antennas at transmitter provides incredible amount of spatial degrees of freedom (DoF), which enables serving many users simultaneously using the same time-frequency resources. Together with enormous transmission bandwidth available in mmWave frequencies, massive MIMO promises an extremely high-speed wireless connection for many users served by the same base station (BS) simultaneously~\cite{Marzetta2010NonCel, Marzetta2013EneSpe, Marzetta2014mMIMO}.

One major challenge with massive MIMO is that the energy consumption and cost of the BS increases very rapidly along with increasing number of antennas. Assuming full-digital precoding at the BS, each transmit antenna is accompanied by a separate D/A converter and a radio frequency (RF) chain, which consists of hardware units to carry the baseband communications signal to mmWave frequency band (e.g., frequency mixer, local oscillator, and power amplifier, etc.). Since the energy consumption and cost of D/A converters increase rapidly along with number of resolution bits (or, equivalently quantization levels), use of massive amount of D/A converters with high resolution is prohibitively undesired \cite{Durisi2017QuaPre, Swindlehurst2017AnaOne, Murmann2018ADC}.

Low-resolution data converters have therefore been considered very favorable in massive MIMO to cut down the cost and energy budget required. In particular, one-bit quantizers are studied extensively in the literature in terms of respective sum spectral efficiency~\cite{Heath2015CapAna, Heath2017UplPer, Heath2018OneBit}. The optimal precoder design is, however, considered in very few studies. In~\cite{Durisi2017QuaPre}, a nonlinear precoder design strategy is considered where computationally complex convex optimization methods (e.g., semidefinite relaxation, sphere precoder, etc.) are employed to compute optimal precoder. A one-bit ZF precoder is proposed by \cite{Swindlehurst2017AnaOne}, which is based on a comparison of the output of the proposed precoder with that of the conventional ZF precoder assuming that unprecoded data symbols are quadrature phase shift keying (QPSK) modulated. In much of the related literature, the conventional RZF and ZF precoders are employed, which are not further optimized taking into account the nonlinear quantization of low-resolution data converters \cite{Xu2017UseLoa, Xu2018OptMul}.

This paper proposes a novel precoder design for a multiuser massive MIMO in mmWave downlink communications, where transmitter employs one-bit D/A converters to come up with a cost- and energy-efficient architecture. In particular, the precoder vector for each user is optimized \textit{individually} by minimizing \textit{energy leakage} into the undesired users. Towards this end, we employ SLNR as an optimization measure to penalty for any energy leakages~\cite{Sayed2007ActAnt, Sayed2007LeaBas, Aissa2011LeaBas}. The proposed design also takes into account any \textit{impairment} due to the nonlinear one-bit quantization. We show that the well-known RZF precoder is indeed a particular version of the proposed SLNR-based precoder when \textit{the quantization impairments are totally ignored}. The numerical results verify that the proposed SLNR-based precoder is significantly superior to both RZF and ZF precoders. To the best of our knowledge, the proposed design is the first to employ SLNR as the optimization measure for multiuser massive MIMO in mmWave communications, where the impairments of one-bit quantization are also incorporated.   

The rest of the paper is organized as follows. Section~\ref{sec:system} introduces the system model under consideration. The linear precoding scheme and linear approximation to one-bit quantization is presented in Section~\ref{sec:linear_precoding}, while the novel SLNR-based precoder design is described in Section~\ref{sec:precoder_design}. Numerical results regarding performance evaluation of the proposed precoder are given in Section~\ref{sec:results}, and the paper concludes with some remarks in Section~\ref{sec:conclusion}.

\textit{Notation:} $\mathbb{E}_X \left\lbrace \cdot \right\rbrace$ denotes ensemble average over the random variable $X$. $\|\cdot\|^2$ and $\|\cdot\|_{\rm F}$ are the Euclidean and Frobenious norm operators, respectively. $[\textbf{X}]_{ii}$ and $[\textbf{x}]_{i}$ represent $i$th diagonal entry of matrix $\textbf{X}$ and $i$th element of vector $\textbf{x}$, respectively. ${\rm diag}(\textbf{X})$ (${\rm diag}(\textbf{x})$) produces a vector (diagonal matrix) consisting of diagonal entries of matrix (entries of vector) $\textbf{X}$ ($\textbf{x}$). $\textbf{I}_{K}$ is the identity matrix of size $K{\times}K$, and $\textbf{0}_{K}$ is the all-zero vector of length $K$. $\mathbb{C}$ denotes complex numbers with real and imaginary parts denoted by $\mathcal{R}$ and $\mathcal{I}$, respectively. $\mathcal{CN}(\textbf{0}_K,\textbf{C})$ stands for the complex Gaussian distribution with mean $\textbf{0}_K$ and covariance $\textbf{C}$.   . 

\section{System Model}\label{sec:system}
\begin{figure}[!t]
	\centering
	\hspace*{-0.0in}
	\includegraphics[width=0.49\textwidth]{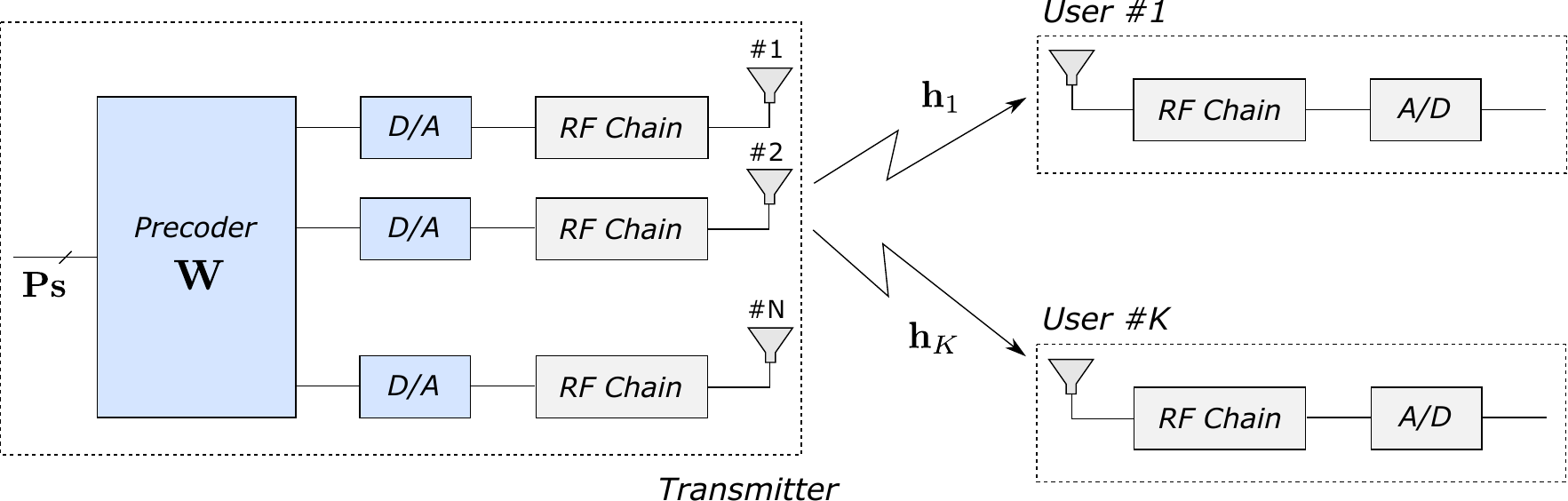}
	\label{fig:antenna_LMS}\vspace{-0.0in}
	\caption{System model for downlink transmission in multiuser mmWave communications with massive MIMO employing one-bit D/A converters.}
	\label{fig:setting}
\end{figure}

We consider a massive multiuser MIMO downlink transmission scenario as shown in Fig.~\ref{fig:setting}, where a BS equipped with $N$ antennas is communicating with $K$ users simultaneously, each of which has a single antenna. We assume full-digital transmitter architecture such that each transmit antenna is accompanied by an RF chain (e.g., frequency mixer, local oscillator, and power amplifier, etc.) and a D/A converter. Considering massive amount of transmit antennas, we assume one-bit quantization for the D/A converters to come up with a cost- and energy-efficient transmitter architecture, whereas the single analog-to-digital (A/D) converter at any of the users has infinite resolution. 

The aggregate received signal $\textbf{y} \,{\in}\, \mathbb{C}^{K{\times}1}$ under the considered scenario is given as
\begin{align} \label{eq:observations}
    \textbf{y} = \textbf{H}^{\rm H} \mathcal{Q}(\textbf{x}) + \textbf{n},
\end{align}
where $\textbf{H} \,{\in}\, \mathbb{C}^{N{\times}K}$ is the aggregate channel matrix, $\mathcal{Q}(\cdot)$ is the one-bit quantization operation, $\textbf{x}\,{\in}\, \mathbb{C}^{K{\times}1}$ is the precoded data, and $\textbf{n}\,{\in}\, \mathbb{C}^{K{\times}1}$ is the circularly symmetric additive white Gaussian noise with the distribution $\mathcal{CN}(\textbf{0}_K,\sigma_{\rm n}^2 \textbf{I}_K)$. 

Considering characteristics of mmWave communications, we assume a correlated channel model for which the channel vector of the $k$th user is given as 
\begin{align} \label{eq:channel}
\textbf{h}_k = \sum\limits_{\ell=1}^{L} \alpha_{\ell} \, \textbf{a}(\theta_{\ell}) ,
\end{align} 
where $L$ is the number of multipaths, $\alpha_{\ell}$ is the complex path gain following the standard complex Gaussian distribution with $\mathcal{CN}(0,\sigma_{\alpha}^2)$, $\theta_{\ell}$ is the angle of departure (AoD), and $\textbf{a}(\theta_{\ell})$ is the array steering vector. Assuming uniform linear array (ULA), the array steering vector is given as  
\begin{align}
\Big[ \textbf{a}(\theta_{\ell}) \Big]_i =  \frac{1}{\sqrt{M}} {\rm exp} \left\lbrace {-}j2\pi \frac{d}{\lambda} \left( i{-}1\right) \cos\left( \theta_{\ell} \right) \right\rbrace,
\end{align}
for $i \,{=}\, 1,\dots,M$, where $d$ is the antenna element spacing along ULA, and $\lambda$ is the wavelength of the carrier frequency. Considering the statistical features of mmWave propagation channels summarized in \cite{Rappaport2017OveMil}, we assume that AoD for each multipath is distributed following Laplace (i.e., double-exponential) distribution with the angular spread $\sigma_{\rm AS}^2$. Note that, the particular user channel $\textbf{h}_k$ in \eqref{eq:channel} is related to the aggregate channel matrix in \eqref{eq:observations} through $\textbf{H} \,{=}\, \left[\textbf{h}_1 \, \textbf{h}_2 \dots \textbf{h}_K \right]$.      

\section{Linear Precoding and Quantization}\label{sec:linear_precoding}
In this section, we briefly introduce the linear precoding scheme under consideration, and describe the linear approximation of the one-bit quantization operation based on the linear precoding scheme.   

\subsection{Linear Precoding}
We assume a linear precoder to get rid of high computational complexity associated with the optimal nonlinear precoder design alternatives~\cite{Durisi2017QuaPre}. The precoder is represented by $\textbf{W}\,{\in}\, \mathbb{C}^{N{\times}K}$, and yields
\begin{align}\label{eq:precoded}
    \textbf{x} = \textbf{W} \textbf{P} \textbf{s},
\end{align}
where $\textbf{P} \,{=}\, {\rm diag} \left(p_1,\dots,p_K\right)$ is the power allocation matrix with $p_k$ being the power allocation coefficient for the $k$th user, and $\textbf{s} \,{\in}\, \mathbb{C}^{K{\times}1}$ is the vector of user messages with the $k$th element $s_k$ representing the $k$th user message, and $\mathbb{E}\left\lbrace \textbf{s} \textbf{s}^{\rm H} \right\rbrace \,{=}\, \sigma_{\rm s}^2 \textbf{I}_{K}$. 

In this work, we adopt a power allocation policy where each user has equal power allocation after precoding~\cite{Heath2017ExpSpa}. Assuming that $\textbf{w}_k$ is the precoder for the $k$th user such that $\textbf{W} \,{=}\, \left[\textbf{w}_1 \, \textbf{w}_2 \dots \textbf{w}_K \right]$, \eqref{eq:precoded} can be decomposed as $\sum_k p_k s_k \textbf{w}_k$, and we therefore have $p_k \,{=}\, \sqrt{\mathsf{P_{TX}}/K \sigma_{\rm s}^2} \left(\|\textbf{w}_k\|\right)^{{-}1}$ with $\mathsf{P_{TX}}$ being the total transmit power. The power allocation coefficients can be jointly represented as
\begin{align}\label{eq:power_allocation}
    \textbf{P} \,{=}\, \sqrt{\frac{\mathsf{P_{TX}}}{K \sigma_{\rm s}^2}}
    \Big[ {\rm diag} \left( \textbf{W}^{\rm H} \textbf{W}\right) \Big]^{-\frac 12},
\end{align}
and the transmit signal-to-noise ratio (SNR) is accordingly given as $\rho \,{=}\, \frac{\mathsf{P_{TX}}}{\sigma_{\rm n}^2}$. 

\subsection{Linear Approximation of One-Bit Quantization}
In order to be able to analyze user rates associated with different precoding schemes, we need to express the nonlinear quantization operation $\mathcal{Q}(\cdot)$ as a linear relation. Assuming that $\textbf{s}$ is complex Gaussian, the quantized data $\textbf{x}_q \,{=}\, \mathcal{Q}(\textbf{x})$ after precoding can be decomposed into a linear function of the precoded data $\textbf{x}$ and a distortion term, according to Bussgang's theorem~\cite{Bussgang1952Cross}. The respective linear approximation is given as
\begin{align}\label{eq:bussgang_0}
    \textbf{x}_q = \mathcal{Q}\left(\textbf{x}\right) = \textbf{A} \textbf{x} + \textbf{q},
\end{align}
where $\textbf{A} \,{\in}\, \mathbb{C}^{K{\times}K}$ is the weight matrix representing the linear quantization operator, and $\textbf{q} \,{\in}\, \mathbb{C}^{K{\times}1}$ is the quantization distortion, which is uncorrelated to the precoded data \textbf{x} as the input to the quantizer. 

In order to have $\textbf{x}_q$ statistically equivalent to $\textbf{x}$, the weight matrix $\textbf{A}$ can be designed based on minimum mean square error (MMSE) criterion as follows
\begin{align}\label{eq:bussgang_1}
    \textbf{A} \textbf{C}_{\textbf{xx}} = \textbf{C}_{\textbf{x}_q \textbf{x}},
\end{align}
where $\textbf{C}_{\textbf{xx}} \,{=}\, \mathbb{E}\left\lbrace \textbf{x} \textbf{x}^{\rm H}\right\rbrace$ is the autocorrelation matrix of unquantized data $\textbf{x}$, and $\textbf{C}_{\textbf{x}_q \textbf{x}} \,{=}\, \mathbb{E} \left\lbrace \textbf{x}_q \textbf{x}^{\rm H}\right\rbrace$ is the cross-correlation matrix of the quantized and unquantized data. Employing \eqref{eq:precoded}, $\textbf{C}_{\textbf{xx}}$ is obtained as follows
\begin{align}\label{eq:autocor_x}
    \textbf{C}_{\textbf{xx}} = \frac{\mathsf{P_{TX}}}{K} \, \textbf{W} 
    \Big[ {\rm diag} \left( \textbf{W}^{\rm H} \textbf{W}\right) \Big]^{-1} \textbf{W}^{\rm H}.
\end{align}
In addition, the cross-correlation matrix for one-bit quantization scheme is given as~\cite{Papoulis2002ProRan, Neri1994EstAut}
\begin{align}\label{eq:crosscorr}
    \textbf{C}_{\textbf{x}_q \textbf{x}} = \sqrt{\frac{2}{\pi}} \Big[ {\rm diag} \left( \textbf{C}_{\textbf{xx}} \right) \Big]^{-\frac{1}{2}} \textbf{C}_{\textbf{xx}}, 
\end{align}
and \eqref{eq:bussgang_1} accordingly becomes
\begin{align}\label{eq:bussgang_2}
    \textbf{A} \textbf{W} \textbf{P}^2 \textbf{W}^{\rm H} = \sqrt{\frac{2}{\pi\sigma_{\rm s}^2}} \Big[ {\rm diag} \left( \textbf{W} \textbf{P}^2 \textbf{W}^{\rm H} \right) \Big]^{-\frac{1}{2}} \textbf{W} \textbf{P}^2 \textbf{W}^{\rm H}.
\end{align}
Note that, since columns of the precoding matrix $\textbf{W}$ corresponds to different users, $\textbf{W}$ has full column rank, and $\textbf{W}^{\rm H} \textbf{W}$ is invertible~\cite{Swindlehurst2017AnaOne}. When we post-multiply both sides of \eqref{eq:bussgang_2} by $\textbf{W}$, we obtain a post-multiplication factor $\textbf{P}^2 \textbf{W}^{\rm H}\textbf{W}$ common at both sides, which is invertible (since $\textbf{P}$ is diagonal with all non-zero diagonal entries), and hence cancels in \eqref{eq:bussgang_2} to yield
\begin{align}\label{eq:bussgang_3}
    \textbf{A} \textbf{W} = \sqrt{\frac{2}{\pi\sigma_{\rm s}^2}} \Big[ {\rm diag} \left( \textbf{W} \textbf{P}^2 \textbf{W}^{\rm H} \right) \Big]^{-\frac{1}{2}} \textbf{W}.
\end{align}  
Note that although a unique expression for $\textbf{A}$ cannot be obtained in \eqref{eq:bussgang_3} since $\textbf{W}\textbf{W}^{\rm H}$ is rank-deficient with $K \,{\ll}\, N$, one solution can be considered by employing \eqref{eq:power_allocation} as 
\begin{align}\label{eq:bussgang_3}
    \textbf{A} = \sqrt{\frac{2 K}{\pi\mathsf{P_{TX}}}}
    \left[ {\rm diag} \left( \textbf{W} \Big[ {\rm diag} \left( \textbf{W}^{\rm H} \textbf{W}\right) \Big]^{-1} \textbf{W}^{\rm H} \right) \right]^{-\frac{1}{2}} \!\!\!,
\end{align}
which is employed in Section~\ref{sec:precoder_design} during the precoder design. 

The term $\textbf{q}$ in \eqref{eq:bussgang_1} representing the quantization distortion is zero mean with the autocorrelation matrix given as 
\begin{align}\label{eq:autocor_q}
    \textbf{C}_{\textbf{qq}} = \textbf{C}_{\textbf{x}_q\textbf{x}_q} - \textbf{A} \textbf{C}_{\textbf{xx}} \textbf{A}^{\rm H}, 
\end{align}
where $\textbf{C}_{\textbf{x}_q\textbf{x}_q}$ is the autocorrelation matrix of the quantized data $\textbf{x}_q$, and is given in terms of the autocorrelation $\textbf{C}_{\textbf{xx}}$ of the unquantized data $\textbf{x}$ through \textit{arcsin law} as follows~\cite{Neri1994EstAut, Papoulis2002ProRan}
\begin{align}\label{eq:autocor_xq}
    &\textbf{C}_{\textbf{x}_q\textbf{x}_q} = \frac{2}{\pi} \bigg[ \arcsin \left( 
    \big[ {\rm diag} \left(\textbf{C}_{\textbf{xx}}\right) \big]^{{-}\frac{1}{2}} 
    \mathcal{R}\left( \textbf{C}_{\textbf{xx}} \right)
    \big[ {\rm diag} \left(\textbf{C}_{\textbf{xx}}\right) \big]^{{-}\frac{1}{2}}\right) \nonumber \\ & \!\!\!+ j \arcsin \left( 
    \big[ {\rm diag} \left(\textbf{C}_{\textbf{xx}}\right) \big]^{{-}\frac{1}{2}} 
    \mathcal{I}\left( \textbf{C}_{\textbf{xx}} \right)
    \big[ {\rm diag} \left(\textbf{C}_{\textbf{xx}}\right) \big]^{{-}\frac{1}{2}}\right) \bigg], 
\end{align}
which can be computed using \eqref{eq:autocor_x}.

The linear approximation in \eqref{eq:bussgang_0} is different from the additive quantization noise model (AQNM) \cite{Erkip2015LowPow, Rangan2007RobPre}, which ignores any correlation between the entries of the quantization distortion (noise) $\textbf{q}$. In contrast, our approximation does not make any particular assumption on the original quantization distortion, and the respective autocorrelation matrix $\textbf{C}_\textbf{qq}$ in ~\eqref{eq:autocor_q} is hence not necessarily diagonal. Note that, one immediate impact of ignoring spatial correlation between the entries of quantization distortion as in AQNM is shown in \cite{Zhang2017OnTra} to overestimate the user rates associated with Gaussian inputs.

\section{SLNR Based Precoder Design}\label{sec:precoder_design}

In this section, we describe the SLNR-based linear precoder design for the multiuser massive MIMO downlink with one-bit quantization. 
\subsection{SINR and SLNR Definitions, and User Sum Rates}

The received signal at the $k$th user can be written by employing \eqref{eq:observations}, \eqref{eq:precoded}, and \eqref{eq:bussgang_0} as follows
\begin{align}\label{eq:observation_k}
    y_k = 
    \underbrace{ \vphantom{\sum\nolimits_{i \neq k}} \textbf{h}_k^{\rm H} \textbf{A} \textbf{w}_k p_k s_k}_{\text{desired signal}} 
    + \underbrace{\sum\nolimits_{i \neq k} \textbf{h}_k^{\rm H} \textbf{A} \textbf{w}_i p_i s_i}_{\text{multiuser interference}} 
    + \underbrace{ \vphantom{\sum\nolimits_{i \neq k}} \textbf{h}_k^{\rm H} \textbf{q}}_{\text{distortion}} + \, n_k,
\end{align}
where $n_k$ is the $k$th entry of the observation noise \textbf{n} of \eqref{eq:observations}. The respective instantaneous SINR (i.e., for a given user channel realization) of the $k$th user can be given as follows
\begin{align}\label{eq:sinr_k}
    \mathsf{SINR}_k = \frac{ p_k^2 
    \textbf{w}_k^{\rm H} \textbf{A}^{\rm H} \textbf{h}_k \textbf{h}_k^{\rm H} \textbf{A} \textbf{w}_k}
    {\sum\nolimits_{i \neq k} p_i^2 
    \textbf{w}_i^{\rm H} \textbf{A}^{\rm H} \textbf{h}_k \textbf{h}_k^{\rm H} \textbf{A} \textbf{w}_i + \textbf{h}_k^{\rm H} \textbf{C}_{\textbf{qq}} \textbf{h}_k + \sigma^2_{\rm n}}.
\end{align}
Note that although the quantization distortion $\textbf{q}$ in \eqref{eq:bussgang_0} is not necessarily Gaussian, a lower bound for the \textit{ergodic} user rates can be obtained by assuming $\textbf{q}$ to be zero-mean Gaussian with the same autocorrelation $\textbf{C}_{\textbf{qq}}$ (since Gaussian noise makes the mutual information minimum) as follows~\cite{Nossek2012CapLow}
\begin{align}\label{eq:sumrate}
    R = \sum_{k=1}^K\mathbb{E}_\textbf{H} \big\lbrace \log_2 \left( 1 + \mathsf{SINR}_k \right) \big\rbrace,
\end{align}
which is normalized by the transmission bandwidth.

In order to maximize the ergodic sum rate, the set of precoders $\textbf{w}_1,\dots,\textbf{w}_K$ have to be optimized jointly considering all the users based on \eqref{eq:sumrate}. Note that SINR of $k$th user is a function of not only the $k$th user's precoder $\textbf{w}_k$, but also the other \textit{interfering} users' precoders $\textbf{w}_i$'s with $i \,{\neq}\, k$. As a result, optimization of $\textbf{w}_k$ needs to be performed by taking into account SINR expressions of all the users through the relation \eqref{eq:sumrate}, which requires the optimization of all the precoders \textit{jointly}. This, in turn, results in non-convex optimization problems which are usually hard to be solved.

In order to get rid of high computational complexity of the ergodic sum rate based optimization for precoder design, we consider the following SLNR expression for the $k$th user
\begin{align}\label{eq:slnr_1}
    \mathsf{SLNR}_k = \frac{ p_k^2 
    \textbf{w}_k^{\rm H} \textbf{A}^{\rm H} \textbf{h}_k \textbf{h}_k^{\rm H} \textbf{A} \textbf{w}_k}
    {\sum\nolimits_{i \neq k} p_k^2 
    \textbf{w}_k^{\rm H} \textbf{A}^{\rm H} \textbf{h}_i \textbf{h}_i^{\rm H} \textbf{A} \textbf{w}_k + \textbf{h}_k^{\rm H} \textbf{C}_{\textbf{qq}} \textbf{h}_k + \sigma^2_{\rm n}},
\end{align}
which is a function of $\textbf{w}_k$ only. The SLNR expression in \eqref{eq:slnr_1} can be considered as a measure of $\textbf{w}_k$ which penalizes for the energy \textit{leaking} into the other users (i.e., those with the index $i$ such that $i \,{\neq}\, k$) in the form of multiuser interference. As a result, the precoder $\textbf{w}_k$ for the $k$th user can be optimized in terms energy leakage into the other users by considering $k$th user's SLNR only~\cite{Sayed2007ActAnt, Sayed2007LeaBas, Aissa2011LeaBas}. This approach therefore enables optimizing each user's precoder individually, which is computationally more efficient than designing all the precoders jointly as in the case of ergodic sum rate based approach. Note, however, that although the SLNR based precoder design does not necessarily maximize the ergodic user rates, the numerical results in Section~\ref{sec:results} verify its superiority as compared to ZF and RZF.       
\subsection{Precoder Design Maximizing SLNR}

The SLNR expression in \eqref{eq:slnr_1} can be elaborated as follows
\begin{align}\label{eq:slnr_2}
    \mathsf{SLNR}_k = \frac{ \textbf{w}_k^{\rm H} \textbf{A}^{\rm H} \textbf{h}_k \textbf{h}_k^{\rm H} \textbf{A} \textbf{w}_k}
    {\textbf{w}_k^{\rm H} \textbf{A}^{\rm H} \left( \textbf{H} \textbf{H}^{\rm H} \,{-}\, \textbf{h}_k \textbf{h}_k^{\rm H} \right) \textbf{A} \textbf{w}_k + \Tilde{\sigma}^2/p_k^2},
\end{align}
where $\Tilde{\sigma}^2 \,{=}\, \textbf{h}_k^{\rm H} \textbf{C}_{\textbf{qq}} \textbf{h}_k \,{+}\, \sigma^2_{\rm n}$ is the composite distortion-plus-noise term specific to the $k$th user. Employing the power allocation policy in \eqref{eq:power_allocation}, and defining $\Tilde{\textbf{h}}_k \,{=}\, \textbf{A}^{\rm H} \textbf{h}_k$ and $\Tilde{\textbf{H}} \,{=}\, \textbf{A}^{\rm H} \textbf{H}$, we have 
\begin{align}\label{eq:slnr_3}
    \mathsf{SLNR}_k = \frac{ \textbf{w}_k^{\rm H} \Tilde{\textbf{h}}_k \Tilde{\textbf{h}}_k^{\rm H} \textbf{w}_k}
    {\textbf{w}_k^{\rm H} \left( \Tilde{\textbf{H}} \Tilde{\textbf{H}}^{\rm H} {-}\, \Tilde{\textbf{h}}_k \Tilde{\textbf{h}}_k^{\rm H} {+}\, \frac{K \sigma_{\rm s}^2 \Tilde{\sigma}^2}{\mathsf{P_{TX}}} \textbf{I}_N \right) \textbf{w}_k}.
\end{align}
Note that \eqref{eq:slnr_3} is maximized whenever the following modified SLNR is maximized, as well.
\begin{align}\label{eq:slnr_4}
    \overline{\mathsf{SLNR}}_k = \frac{ \textbf{w}_k^{\rm H} \Tilde{\textbf{h}}_k \Tilde{\textbf{h}}_k^{\rm H} \textbf{w}_k}
    {\textbf{w}_k^{\rm H} \left( \Tilde{\textbf{H}} \Tilde{\textbf{H}}^{\rm H} {+}\, \frac{K \sigma_{\rm s}^2 \Tilde{\sigma}^2}{\mathsf{P_{TX}}} \textbf{I}_N \right) \textbf{w}_k},
\end{align}
which is a generalized Rayleigh quotient, and is maximized when $\textbf{w}_k$ satisfies the following condition
\begin{align}\label{eq:rayleigh_quotient}
    \!\!\textbf{w}_k \propto \text{max. eigenvector} \left\lbrace \! \left( \Tilde{\textbf{H}} \Tilde{\textbf{H}}^{\rm H} {+} \frac{K \sigma_{\rm s}^2 \Tilde{\sigma}^2}{\mathsf{P_{TX}}} \textbf{I}_N \! \right)^{\!\!\!{-}1} \!\! \Tilde{\textbf{h}}_k \Tilde{\textbf{h}}_k^{\rm H} \right\rbrace \!.
\end{align}
The closed form solution of $\textbf{w}_k$ satisfying \eqref{eq:rayleigh_quotient} is \cite{Doufexi2012OnEqu, Doufexi2013EquExp, Teh2015EneSpe}
\begin{align}\label{eq:optimal_precoder}
    \textbf{w}_k \,{=} \left( \! \textbf{A}^{\rm H}\textbf{H} \textbf{H}^{\rm H}\textbf{A} {+}\, \frac{K \sigma_{\rm s}^2 }{\mathsf{P_{TX}}} \left( \textbf{h}_k^{\rm H} \textbf{C}_{\textbf{qq}} \textbf{h}_k \,{+}\, \sigma^2_{\rm n} \right) \textbf{I}_N \! \right)^{\!\!\!{-}1} \!\! \textbf{A}^{\rm H} \textbf{h}_k.
\end{align}
Note that the SLNR-based optimal precoder $\textbf{w}_k$ in \eqref{eq:optimal_precoder} is a function of not only the channel matrix $\textbf{H}$, but also the weight matrix $\textbf{A}$ of the linear approximation of the one-bit quantization, and the covariance of the quantization distortion $\textbf{C}_{\textbf{qq}}$. As a result, the proposed SLNR-based precoder minimizes energy-leakage taking into account the impairments coming from (linear representation of) the quantization. Whenever we assume ideal quantization, i.e., $\textbf{A} \,{=}\, \textbf{I}_N$ and $\textbf{C}_{\textbf{qq}} \,{=}\, \textbf{0}_N$, the optimal precoder expression in \eqref{eq:optimal_precoder} becomes
\begin{align}\label{eq:rzf}
    \textbf{w}_k \,{=} \left( \! \textbf{H} \textbf{H}^{\rm H} {+}\, \frac{K \sigma_{\rm s}^2 }{\mathsf{P_{TX}}} \sigma^2_{\rm n} \textbf{I}_N \! \right)^{\!\!\!{-}1} \!\! \textbf{h}_k,
\end{align}
which is the $k$th column of the standard RZF precoder (i.e., precoding vector for the $k$th user). As a result, standard RZF does not take into account the impairments of the quantization through either the weight matrix $\textbf{A}$ or the quantization distortion $\textbf{q}$ described in \eqref{eq:bussgang_0}. In Section~\ref{sec:results}, numerical results verify that SLNR-based precoder in \eqref{eq:optimal_precoder} is superior over the RZF precoder in \eqref{eq:rzf} since it optimizes the energy leakage taking into account quantization impairments (i.e., $\textbf{A}$ and $\textbf{q}$).

We note that the weight matrix $\textbf{A}$ involved in \eqref{eq:optimal_precoder} is a function of all the precoders through the precoder matrix $\textbf{W}$ as shown in \eqref{eq:bussgang_3} through Bussgang's theorem. In addition, the the autocorrelation of the quantization distortion $\textbf{C}_{\textbf{qq}}$ is also a function of $\textbf{W}$ as given by \eqref{eq:autocor_x}, \eqref{eq:autocor_q}, and \eqref{eq:autocor_xq}. As a result, stacking optimal precoder vectors given in \eqref{eq:optimal_precoder} side-by-side to construct $\textbf{W}$, the optimal solution for the SLNR-based precoder design can be expressed as
\begin{align}
    \textbf{W} = f\left(\textbf{W},\textbf{H}\right),
\end{align}
for an adequate function $f(\cdot)$ acting element-wise on its argument matrix. Note that the optimal SLNR-based precoder matrix $\textbf{W}$ turns out to be the multidimensional fixed-point of function $f(\cdot)$, which can be solved through fixed-point iterations as described in Algorithm~\ref{alg:alg}.

\begin{algorithm}
\caption{SLNR-Based Precoder Design}
\begin{algorithmic}[1]
\State Initialize precoder $\textbf{W}_1$ and error tolerance $\varepsilon$
\State $k \gets 1$
\While{ $\|\textbf{W}_k \,{-}\, \textbf{W}_{k{-}1}\|_{\rm F} \,{>}\, \varepsilon$ }
    \State $\textbf{W} \gets \textbf{W}_k$ 
    \State Compute $\textbf{A}$ by \eqref{eq:bussgang_3}, $\textbf{C}_{\textbf{xx}}$ by \eqref{eq:autocor_x}, and $\textbf{C}_{\textbf{x}_q\textbf{x}_q}$ by \eqref{eq:autocor_xq}
    \State Compute $\textbf{C}_{\textbf{qq}}$ by using $\textbf{A}$, $\textbf{C}_{\textbf{xx}}$, and $\textbf{C}_{\textbf{x}_q\textbf{x}_q}$ in \eqref{eq:autocor_q}
    \State Update $\textbf{W}$ by using $\textbf{A}$, $\textbf{C}_{\textbf{qq}}$, and $\textbf{H}$ in \eqref{eq:optimal_precoder} 
    \State $\textbf{W}_{k{+}1} \gets \textbf{W}$ 
    \State $k \gets k + 1$
\EndWhile
\end{algorithmic}\label{alg:alg}
\end{algorithm}

\section{Numerical Results}\label{sec:results}
In this section, we present numerical results based on extensive Monte Carlo simulations, to evaluate the performance of the proposed SLNR-based precoder for multiuser massive MIMO communications in mmWave frequencies with one-bit quantization. Considering the mmWave propagation characteristics \cite{Rappaport2017OveMil}, the simulation parameters are listed Table~\ref{tab:simulation_parameters}. 
\begin{table}[!h]
\renewcommand{\arraystretch}{1.3}
	\caption{Simulation Parameters}
	\label{tab:simulation_parameters}
	\centering
	\begin{tabular}{lc}
		\hline
		Parameter & Value \\
		\hline\hline
		Number of users $(K)$ & $10-100$ \\
		Number of transmit antennas $(N)$ & $100$ \\
		Transmit SNR $(\rho)$ & $\{10,40\}~\text{dB}$ \\
		Number of Multipaths $(L)$ & $5$\\
		AOD distribution & Laplace \\
		Angular spread $(\sigma^2_{\rm AS})$ & $5^\circ$\\
		User angular position & Uniform in $\left[0^\circ,90^\circ\right]$\\
		Number of iterations $(I)$ & $5$ \\
        Average symbol energy $(\sigma^2_{\rm s})$ & $1$ Watts\\
        Variance of complex path gain $(\sigma_{\alpha}^2)$ & 1 \\
		\hline
	\end{tabular}
\end{table}

For comparison purposes, we also provide numerical results associated with the well-known RZF precoder given in \eqref{eq:rzf}, and the ZF precoder given as follows
\begin{align}\label{eq:zf}
    \textbf{W} = \textbf{H} \left( \textbf{H}^{\rm H} \textbf{H} \right)^{{-}1},
\end{align}
which is also employed in the SLNR-based precoder initialization step described in  Algorithm~\ref{alg:alg}.

\begin{figure}[!t]
\centering
\subfloat[SNR$\,{=}\,10~\text{dB}$]
{\includegraphics[width=0.47\textwidth]{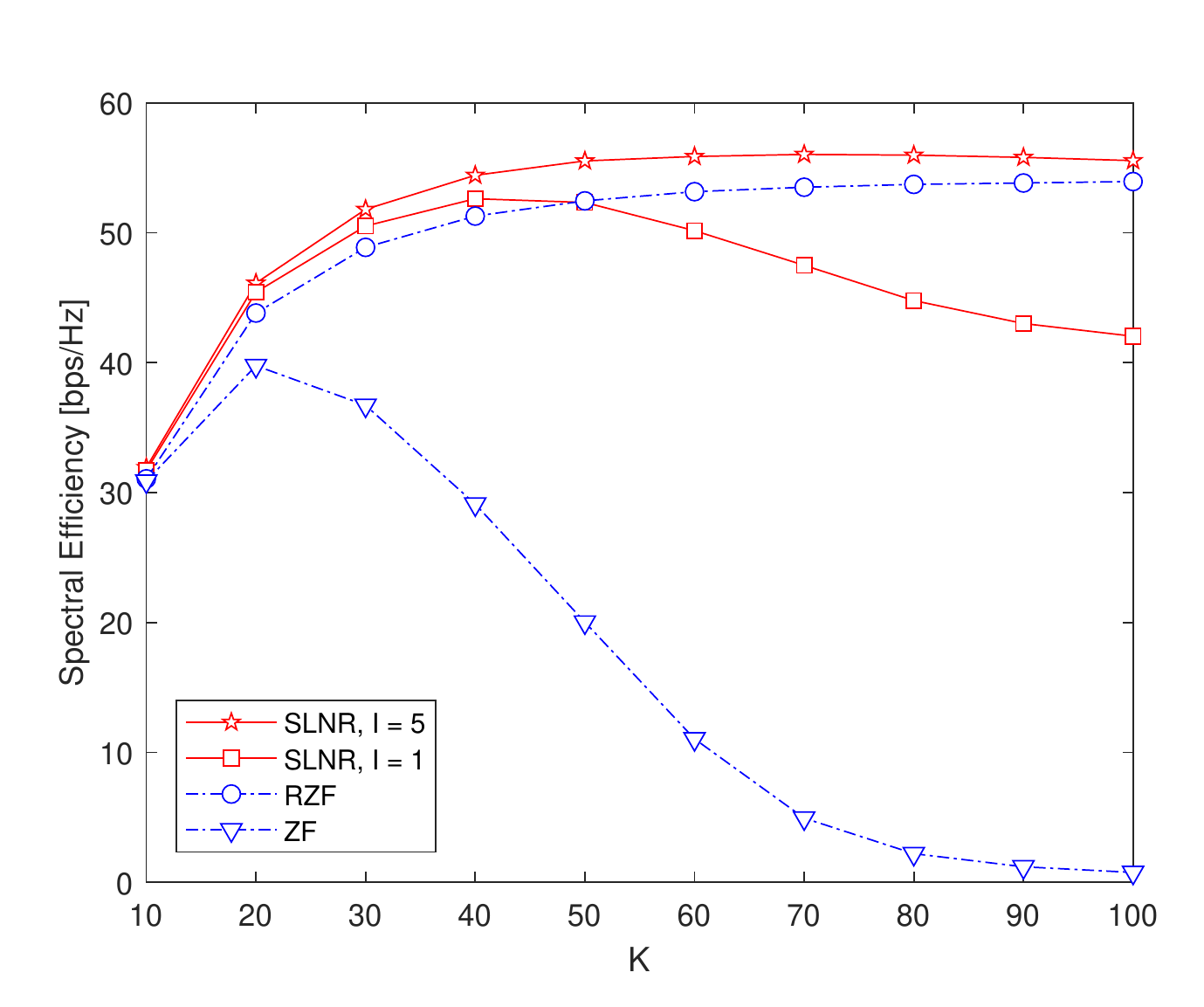}
\label{fig:rates_snr10db_mpc5_ang090}}\\
\vspace{-0.2in}
\subfloat[SNR$\,{=}\,40~\text{dB}$]
{\includegraphics[width=0.47\textwidth]{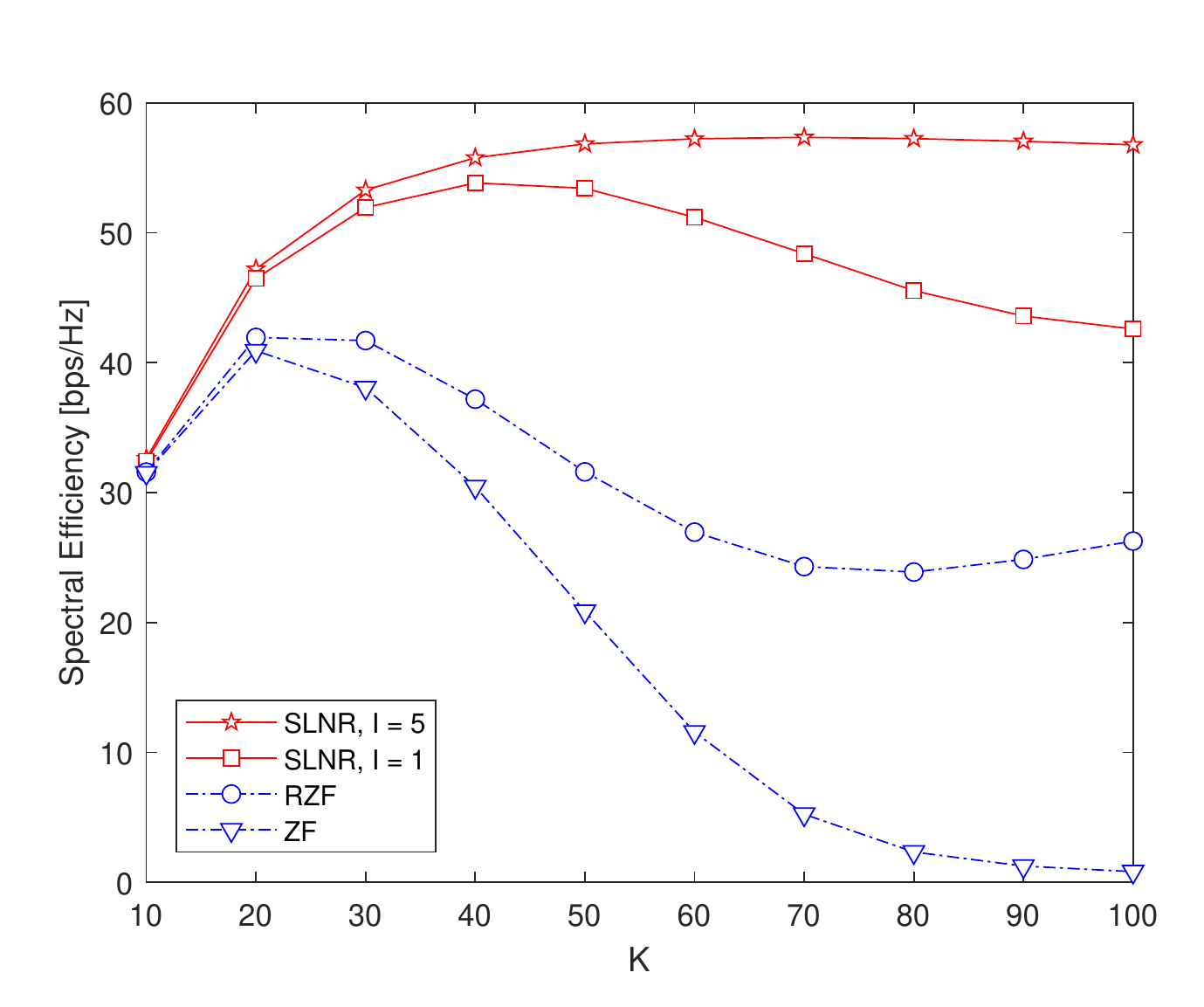}
\label{fig:rates_snr40db_mpc5_ang090}}
\caption{Sum spectral efficiency for varying number of users $K$ and iterations $I$ at SNR of $\{10,40\}~\text{dB}$.}
\label{fig:rates_mpc5_ang090}
\end{figure}

In Fig.~\ref{fig:rates_mpc5_ang090}, we depict sum spectral efficiency, which is the user sum rates normalized by the transmission signal bandwidth, for SLNR-based precoder together with RZF and ZF precoders against number of users $K \,{\in}\, [10,100]$. Defining $\beta \,{=}\, N/K$ to be the user-loading factor, the presented results cover both the low (i.e., $\beta \,{=}\, 0.1$) and the full (i.e., $\beta \,{=}\, 1$) user-loading scenarios. We observe that the SLNR-based precoder significantly outperforms RZF and ZF at both $10~\text{dB}$ and $40~\text{dB}$ SNR values presented in Fig.~\ref{fig:rates_mpc5_ang090}\subref{fig:rates_snr10db_mpc5_ang090} and Fig.~\ref{fig:rates_mpc5_ang090}\subref{fig:rates_snr40db_mpc5_ang090}, respectively. Note that the performance of RZF at the high SNR value of $40~\text{dB}$ exhibits a deteriorating trend along with increasing number of users, which makes it worse than the SLNR-based precoder obtained after even the $1$st iteration only.

Note that RZF follows a similar (but still worse) variation trend in spectral efficiency as compared to SLNR-based precoder when SNR is $10~\text{dB}$, as shown in Fig.~\ref{fig:rates_mpc5_ang090}\subref{fig:rates_snr10db_mpc5_ang090}. As SNR increases, the composite distortion-plus-noise term $\Tilde{\sigma}^2 \,{=}\, \textbf{h}_k^{\rm H} \textbf{C}_{\textbf{qq}} \textbf{h}_k \,{+}\, \sigma^2_{\rm n}$ in \eqref{eq:optimal_precoder} is dominated by the distortion part $\textbf{h}_k^{\rm H} \textbf{C}_{\textbf{qq}} \textbf{h}_k$ (since $\sigma^2_{\rm n} \,{\rightarrow}\, 0$), which is totally ignored by RZF (and also its particular version ZF). As a result, the sum spectral efficiency performance of RZF rapidly deteriorates with increasing number of users when SNR is increased up to $40~\text{dB}$, as shown in Fig.~\ref{fig:rates_mpc5_ang090}\subref{fig:rates_snr40db_mpc5_ang090}. On the other hand, the SLNR-based precoder still remains the same sum spectral efficiency along with increasing number of users without any degradation, since it takes into account the quantization distortion $\textbf{q}$ as well as the weight matrix $\textbf{A}$ in \eqref{eq:bussgang_0}. 

\begin{figure}[!t]
	\centering
	\hspace*{-0.0in}
	\includegraphics[width=0.47\textwidth]{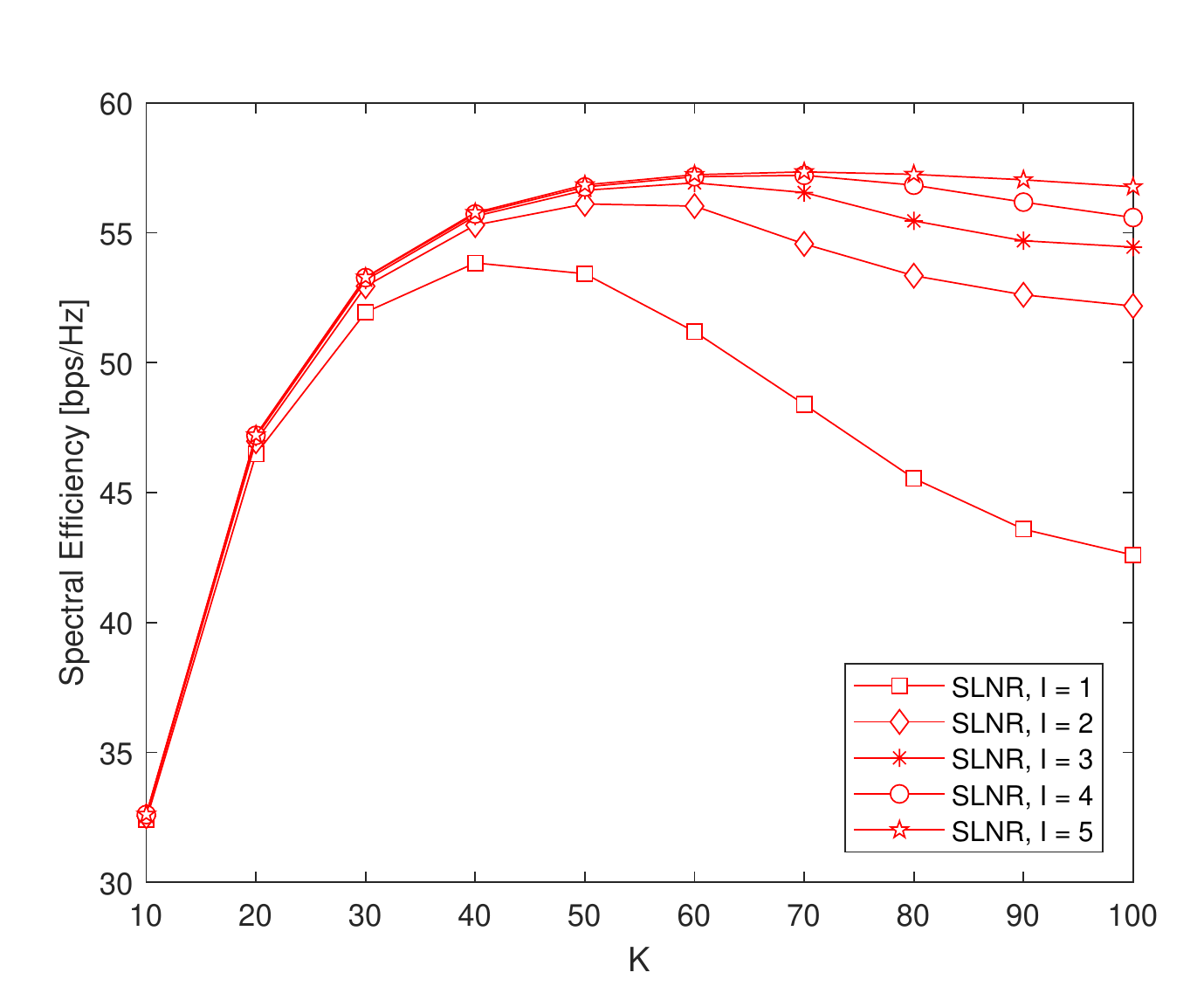}
	\label{fig:antenna_LMS}\vspace{-0.0in}
	\caption{Convergence of sum spectral efficiency of SLNR-based precoder along with varying number of users $K$ and iterations $I$ at $10~$dB of SNR.}
	\label{fig:convergence_snr10db_mpc5_ang090}
\end{figure}

We also observe that the number of iterations required for the SLNR-based precoder to converge increases as the number of users $K$ gets larger regardless of the particular SNR value, as shown in Fig.~\ref{fig:rates_mpc5_ang090}. This is actually consistent with the rationale behind the SLNR-based precoder as it is designed to optimize the energy leakage into the other users, which naturally becomes more complicated as the number users to take into account increases. The convergence behaviour of the SLNR-based precoder is explicitly depicted in Fig.~\ref{fig:convergence_snr10db_mpc5_ang090} for $10~\text{dB}$ and $5$ iterations in total, which underscores this conclusion.     

\begin{figure}[!t]
\centering
\subfloat[SNR$\,{=}\,10~\text{dB}$]
{\includegraphics[width=0.47\textwidth]{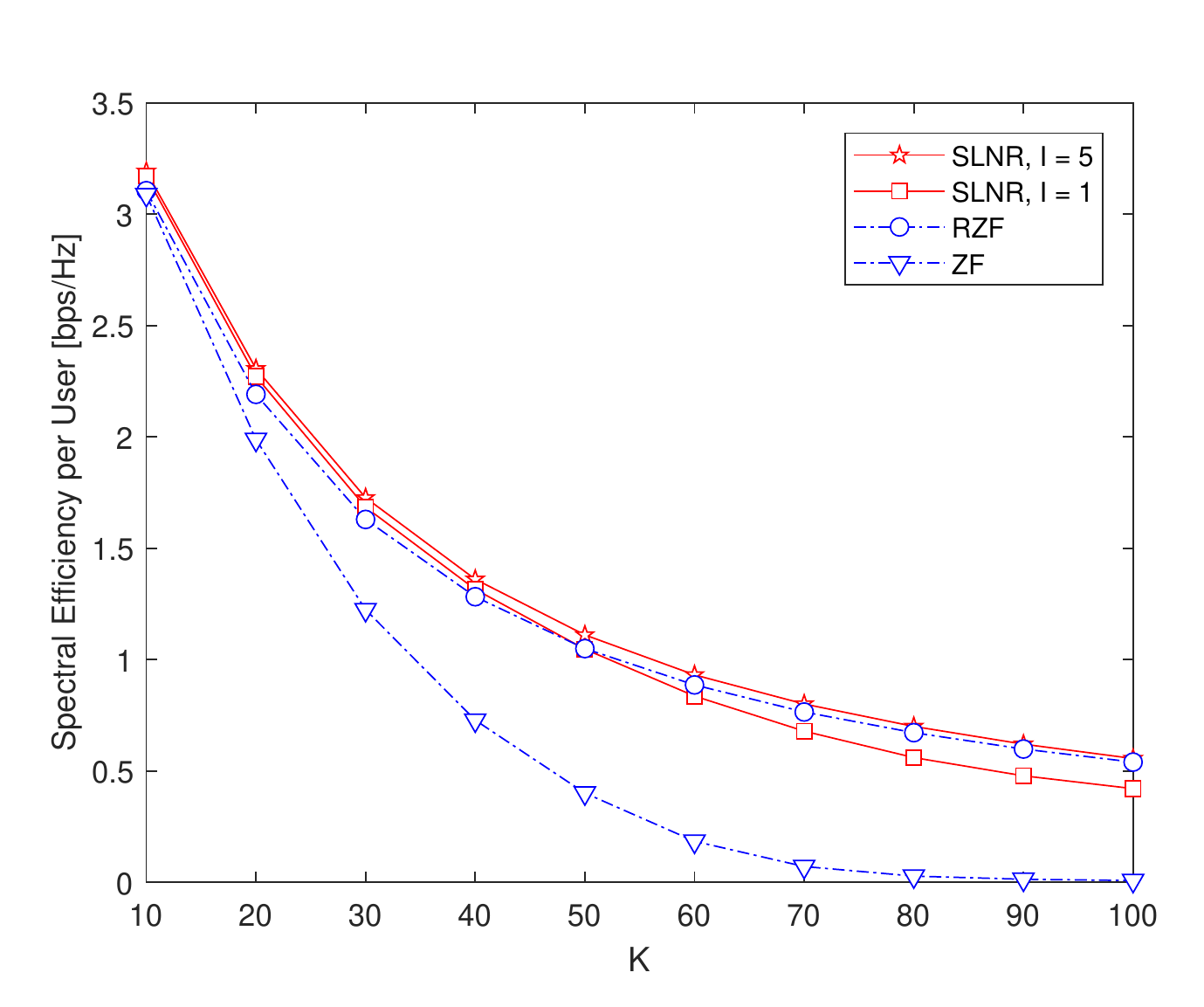}
\label{fig:userload_snr10db_mpc5_ang090}}\\
\vspace{-0.2in}
\subfloat[SNR$\,{=}\,40~\text{dB}$]
{\includegraphics[width=0.47\textwidth]{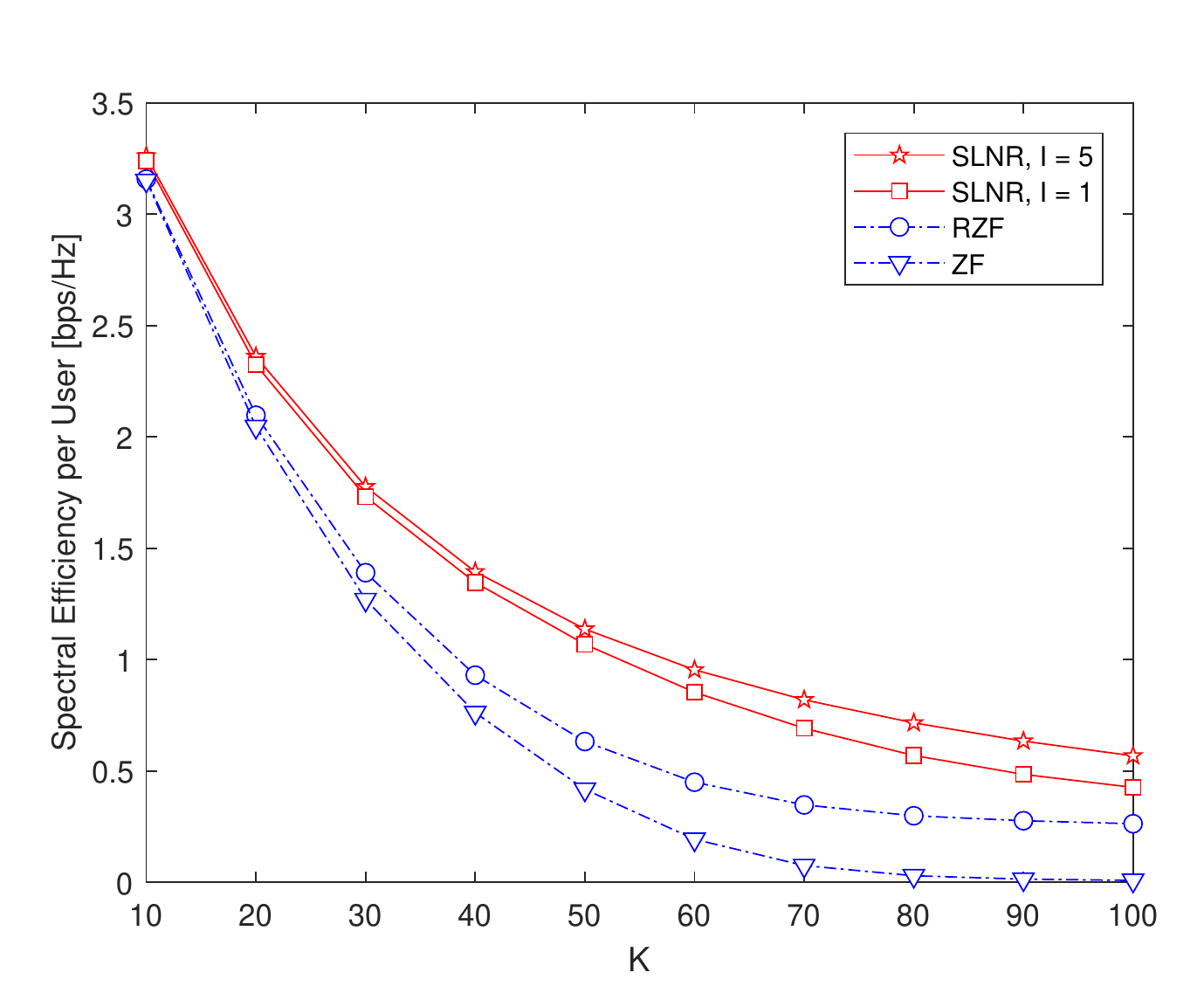}
\label{fig:userload_snr40db_mpc5_ang090}}
\caption{Spectral efficiency per user for varying number of users $K$ and iterations $I$ at SNR of $\{10,40\}~\text{dB}$.}
\label{fig:userload_mpc5_ang090}
\end{figure}

In Fig.~\ref{fig:userload_mpc5_ang090}, we depict numerical results for the spectral efficiency per user. As the number of users $K$ increases, the spectral efficiency per user decreases, as expected. Similar to the sum spectral efficiency results of Fig.~\ref{fig:rates_mpc5_ang090}, we observe that the SLNR-based precoding utilizes the spectral resources better than either RZF or ZF. As before, the performance gap between the SLNR-based precoder and RZF widens significantly at high SNR (i.e., $40~\text{dB}$), since RZF ignores the impact of quantization distortion $\textbf{q}$, which becomes dominant at high SNR, while the SLR-based precoder is optimized based on both $\textbf{q}$ and the weight matrix $\textbf{A}$ associated with one-bit quantization.      

\section{Conclusion}\label{sec:conclusion}
In this work, we consider a multiuser massive MIMO communications in mmWave frequency bands, where the transmitter employs one-bit D/A converters to end up with a cost- and energy-efficient architecture. We propose a precoder design based on the minimization of the energy leakage into the users other than the desired one. Although the standard RZF and ZF precoders ignore any distortion introduced by quantization through D/A converters, the proposed design takes into account such quantization impairments. Numerical results verify the superiority of the proposed precoder design over RZF and ZF.     

\bibliographystyle{IEEEtran}
\bibliography{IEEEabrv,references}

\end{document}